\documentclass[pre,aps,groupaddress,showpacs,prd,twocolumn]{revtex4-1}
\usepackage{epsfig,amsmath,graphicx,amssymb,overpic}

\usepackage{graphicx}
\usepackage{dcolumn}
\usepackage{bm}
\usepackage{graphicx}
\usepackage{subfigure}

\def\be{\begin{equation}}
\def\ee{\end{equation}}
\def\bee{\begin{eqnarray}}
\def\ene{\end{eqnarray}}
\def\bes{\begin{subequations}}
\def\ees{\end{subequations}}

\def\PT{\mathcal{PT}}

\begin{document}

\title{Dynamics of higher-order rational solitons for the nonlocal nonlinear
Schr\"odinger equation with the self-induced parity-time-symmetric potential}

\author{Xiao-Yong Wen$^{1,2}$}
\author{Zhenya Yan$^1$}\email{Corresponding author. {\it E-mail address}: zyyan@mmrc.iss.ac.cn}
\author{Yunqing Yang$^{3}$}
\affiliation{\vspace{0.1in} $^1$Key Laboratory of Mathematics Mechanization, Institute
of Systems Science, AMSS, \\ Chinese Academy of Sciences, Beijing
100190, China\\ $^2$Department of Mathematics, School of Applied Science, Beijing Information
    Science and Technology University, Beijing  100192, China \\
$^{3}$School of Mathematics, Physics and Information Science, Zhejiang Ocean University, Zhoushan, Zhejiang  316022, China\vspace{0.1in}}

\date{\vspace{0.1in} 31 Jan. 2016,  Chaos {\bf 26}, 063123 (2016)}

\begin{abstract}  The integrable nonlocal nonlinear Schr\"odinger (NNLS) equation with the self-induced parity-time-symmetric potential [Phys. Rev. Lett. 110 (2013) 064105] is investigated, which is an integrable extension of the standard NLS equation. Its novel higher-order rational solitons are found using the nonlocal version of the generalized perturbation $(1, N-1)$-fold Darboux transformation. These rational solitons illustrate abundant wave structures for the distinct choices of parameters (e.g., the strong and weak interactions of bright and dark rational solitons). Moreover, we also explore the dynamical behaviors of these higher-order rational solitons with some small noises on the basis of numerical simulations.

  \end{abstract}


\maketitle
\baselineskip=12pt

\textbf{The study of nonlinear waves and soliton theory has become a more and more significant subject in many branches of nonlinear science. The fundamental  solitons (e.g., bright and dark solitons) are usually expressed in terms of fractional formals of exponential functions. They more quickly  tend to some constants than the localized rational solutions as the variables approach to infinity. The localized rational solutions of nonlinear wave equations admit the special properties, one of which is that they may have the finite critical points (e.g., rational rogue waves) or infinite many critical points (e.g., rational solitons). Recently, a new integrable nonlocal nonlinear Schr\"odinger (NNLS) equation  with the self-induced parity-time-($\PT$-) symmetric potential was presented [Phys. Rev. Lett. 110 (2013) 064105]. In this paper, we present the nonlocal version of the generalized perturbation Darboux transformation to study the novel higher-order rational solitons of the NNLS equation, which exhibit the abundant wave structures for the different choices of parameters (e.g., interactions of bright and dark rational solitons). Moreover, we also investigate their dynamical behaviors with some small noises by using numerical simulations. These results would be useful for understanding the corresponding rational soliton phenomena in the many fields of nonlocal nonlinear dynamical systems such as nonlinear optics, Bose-Einstein condensates, ocean, and  other relevant fields. }

\section{Introduction}

Recently, rogue waves (RWs, a special type of rational solitons), originally occurring in the deep ocean~\cite{org,org2,org3,org4}, attracted more and more theoretical and experimental attention in many other fields such as nonlinear optics~\cite{orw,orw2,orw3}, hydrodynamics~\cite{hd}, Bose-Einstein condensates~\cite{bec1,bec2}, plasma~\cite{pla}, and even finance~\cite{yan,yan2}. RWs are also called freak waves~\cite{fw}, giant waves, great waves, huge waves, ginormous waves, ghost waves, killer waves, etc. The danger of oceanic RWs is due to their sudden appearance and disappearance as `waves appear from nowhere and disappear without a trace'~\cite{nail09, nailpla}. Moreover, the word `rogon' was coined for the RWs if they reappear virtually unaffected in size or shape shortly after their interactions~\cite{yanpla10}, which is similar to `soliton'.

The integrable nonlinear Schr\"odinger (NLS) equation~\cite{nls0,nls1,nls4,nls2,nls3}
\bee \label{nls1}
  iq_t-\frac{1}{2}q_{xx}-\sigma|q|^2q=0, \qquad q\equiv q(x,t)
 \ene
appears in many fields of nonlinear science such as nonlinear optics, the deep ocean, DNA, and Bose-Einstein condensates, where the subscript denotes the partial derivative with respect to the variables and $\sigma=\pm 1$.  It is an importantly nonlinear integrable model admitting explicit first-order RW solution (also called Peregrine's RW solution) for the focusing case $\sigma=1$~\cite{ps} $q_{\rm ps}(x,t)=\left[1-\frac{4(1-2it)}{1+4(x^2+t^2)}\right]e^{-it}$,
which can be regarded as the parameter limit of its breathers~\cite{MA,nail86,nail87,rgre}, and  higher-order RW solutions~\cite{nail,drw01,drw02,drw03,mat3,mat4,guo1}. The intensity, $|q_{\rm ps}|^2$, is localized in both space and time and approaches to one not zero as $x^2+t^2\rightarrow \infty$, which differs from its bright soliton ($\sigma=1$), $q_{\rm b}(x,t)=\beta\, {\rm sech}(\beta x+v t)\exp\left[i\left(\frac{v}{\beta}x+\frac{v^2-\beta^4}{2\beta^2}t\right)\right]\,(\beta\not=0,\, v \in\mathbb{R})$,
 in which $|q_{\rm b}|^2\rightarrow 0$ as $x^2+t^2\rightarrow \infty$. It has been shown that the Peregrine's RW solution has a good agreement with the numerical and experimental results of Eq.~(\ref{nls1})~\cite{orw3}. But the NLS equation with the defocusing case $\sigma=-1$ was shown to possess the singularly rational solutions.

 Recently, a new nonlocal nonlinear Schr\"odinger (NNLS) equation  with the self-induced $\PT$-symmetric potential was presented in the form~\cite{nnls}
 \begin{eqnarray}
     iq_t(x,t)-\frac{1}{2}q_{xx}(x,t)-\sigma q^2(x,t) q^{*}(-x,t)=0, \label{cmkdv}
\end{eqnarray}
where the subscript denotes the partial derivative with respect to the variables, the star stands for the complex conjugation, and $\sigma=\pm 1$ corresponds to the self-focusing case $\sigma=1$ and defocusing case $\sigma=-1$, respectively. Eq.~(\ref{cmkdv}) can be regarded as an {\it integrable extension} of Eq.~(\ref{nls1}) with $q^*(x,t)\to q^*(-x,t)$. Eq.~(\ref{cmkdv}) was still verified to be completely integrable, that is, it admits the Lax pair, infinite conversation laws, etc., but Eq.~(\ref{cmkdv}) and Eq.~(\ref{nls1}) are different. The solitons and breather solutions of Eq.~(\ref{cmkdv}) have been studied~\cite{lm}. The semi-linear operator related to Eq.~(\ref{cmkdv}) ${\mathcal H}=-\frac12\partial_x^2-\sigma q(x,t)q^{*}(-x,t)$ is $\PT$-symmetric for any solutions of Eq.~(\ref{cmkdv}), where the complex $\PT$-symmetric potential $W(x,t)=-\sigma q(x,t)q^{*}(-x,t)$ is regarded to be self-induced and $t$ is a parameter for the linear spectral problem, and the operators $\mathcal{P}$ and $\mathcal{T}$ are defined by $\mathcal{P}$: $x\to -x$ and $\mathcal{T}$: $i\to -i$~\cite{pt}.

We now simply compare the NLS equation (\ref{nls1}) with NNLS equation (\ref{cmkdv}). (i) If $q(-x,t)\equiv q(x,t)$, then Eq.~(\ref{cmkdv}) becomes the NLS equation (\ref{nls1}). That is, if the solutions of NLS equation (\ref{nls1}) are the even functions for $x$, then its solutions must be ones of the NNLS equation Eq.~(\ref{cmkdv}); (ii) If $q(-x,t)\equiv -q(x,t)$, then Eq.~(\ref{cmkdv}) becomes the NLS equation (\ref{nls1}) with $\sigma\to -\sigma$. That is, if the solutions of NLS equation (\ref{nls1}) are the odd functions for $x$, then its solutions must be ones of the NNLS equation (\ref{cmkdv}) with $\sigma\to -\sigma$. For example, the first-order RW solution $q_{\rm ps}(x,t)$ and higher-order RW solutions of NLS equation (\ref{nls1}) also solve Eq.~(\ref{cmkdv}) with  $\sigma=1$. The above-mentioned bright solution $q_{\rm b}(x,t)$ is not an even or odd function for $x$ for the non-zero wave speed parameter $v$, but it is an even function for $x$ for the zero speed $v=0$ such that $q_{\rm b}(x,t)$ with $v=0$ is also a bright soliton of NNLS equation (\ref{cmkdv}) with $\sigma=1$ (see Ref.~\cite{yanaml15} for other special solutions).  To the best of our knowledge, the higher-order rational solitons (which are the neither  even nor odd functions for $x$) and dynamical behaviors of Eq.~(\ref{cmkdv}) were not considered before.

Recently, some power methods have been developed to investigate the higher-order RW solutions of nonlinear wave equations such as the modified and generalized Darboux transformation (DT) ~\cite{nail,drw01,drw02,drw03,mat3,mat4,guo1,guo2,he, yanchaos15}, the Hirota's bilinear method with the $\tau$-function~\cite{yang1,yang2}, the similarity (symmetry) transformation~\cite{yanpla10,yan12,yan12b}, and so on.
Recently, we presented a generalized perturbation $N$-fold Darboux transformation to find higher-order RW solutions of modified NLS equation~\cite{wen1} and generalized integrable coupled AB system~\cite{wen2}, which are both local models.

In this paper, we will extend our previous method used in the local equations~\cite{wen1, wen2} to present the {\it nonlocal version} of the local $(1,N-1)$-fold Darboux transformation in terms of the Taylor series expansion for the parameter and a limit procedure to directly obtain  higher-order rational solitons of the NNLS equation (\ref{cmkdv}). The biggest advantage of our method is to obtain the higher-order rational solitons in terms of determinants without complicated iterations, and the relationships between the multi-rational solitons and the `seed' solutions are established clearly.

\section{The nonlocal nonlinear Schr\"odinger equation}

\subsection{Lax pair and gauge transformation}

To study the novel localized solutions (e.g., regular rational solitons) of  Eq.~(\ref{cmkdv}), we need to present its generalized perturbation
 $(1, N-1)$-fold DT of Eq.~(\ref{cmkdv}) in terms of its Lax pair. We firstly consider the Lax pair (or the linear iso-spectral problems) of Eq.~~(\ref{cmkdv}) in the form~\cite{nnls}
\begin{eqnarray}
\varphi_x=U \varphi, \quad U=\left(\begin{array}{cc}\lambda & q(x,t) \vspace{0.05in} \\  -\sigma q^{*}(-x,t)  &-\lambda \end{array}\right),  \label{lax1}
\end{eqnarray}
\begin{eqnarray}
 \varphi_{t}=V \varphi,\quad V\!=\!\left(\!\begin{array}{cc}
        V_{11}   &  V_{12}  \vspace{0.05in} \\
       V_{21}    &  V_{22}  \end{array} \!\right), \,\,\, \label{lax2}
\end{eqnarray}
with \bee\nonumber\begin{array}{l}
 V_{11}=-i\lambda^2-\frac{i}{2}\sigma q(x,t) q^{*}(-x,t), \vspace{0.05in} \\
  V_{12}=-i \lambda q(x,t) -\frac{i}{2} q_x(x,t),\vspace{0.05in} \\
 V_{21}=i\sigma \lambda q^{*}(-x,t) +\frac{i}{2} \sigma q_x^{*}(-x,t),\vspace{0.05in} \\
  V_{22}=i\lambda^2+\frac{i}{2}\sigma q(x,t) q^{*}(-x,t)
\end{array}\ene
where the star represents the complex conjugation, $\varphi=(\phi,\psi)^T$ (the superscript $T$ denotes the vector transpose) is the vector eigenfunction, $\lambda$ is the spectral parameter,  and $i^2=-1$. It is easy to show that the compatibility condition $\varphi_{xt}=\varphi_{tx}$, that is, zero curvature equation $U_t-V_x+[U, V]=0$, of Lax pair (\ref{lax1})-(\ref{lax2}) just leads to Eq.~(\ref{cmkdv}).

We now consider the gauge (symmetry) transformation~\cite{dt3}
\bee \label{dm}
 \widetilde{\varphi}=T(\lambda)\varphi, \, \widetilde{\varphi}=(\widetilde{\phi},\,  \widetilde{\psi})^T
 \ene
of the Lax pair (\ref{lax1}) and (\ref{lax2}), where $T(\lambda)$ is the unknown $2\times 2$ Darboux matrix, and the new eigenfunction $\widetilde{\varphi}$ satisfies the same Lax pair (\ref{lax1}) and (\ref{lax2}) with the old potential function $q(x,t)$ being replaced by the new one $\widetilde{q}(x,t)$, i.e.,
 \begin{eqnarray}
\widetilde{\varphi}_x=\widetilde{U} \widetilde{\varphi}, \quad \widetilde{U}=U|_{q(x,t)\to \widetilde{q}(x,t),\, q(-x,t)\to \widetilde{q}(-x,t)},  \label{nlax3}
 \qquad
\end{eqnarray}
\bee
 \widetilde{\varphi}_t=\widetilde{V}\widetilde{\varphi},\quad \widetilde{V}=V|_{q(x,t)\to \widetilde{q}(x,t),\, q(-x,t)\to \widetilde{q}(-x,t)}.  \label{nlax4}
\ene

Therefore, on the basis of Eqs.~(\ref{nlax3})-(\ref{nlax4}) we have
\bee \label{t1}
T_{x}+[T,\, \{U,\, \widetilde{U}\}]=0, \qquad
T_{t}+[T,\, \{V,\,  \widetilde{V}\}]=0,
\label{t2}\ene
where we have introduced the generalized bracket $[F, \{G,  \widetilde{G}\}]=FG-\widetilde{G}F$. In particular, the  generalized bracket reduces to Lie bracket $[F, \{G, \widetilde{G}\}]=[F, G]$ for the case $\widetilde{G}=G$. Therefore we have
$\widetilde{U}_t-\widetilde{V}_x+[\widetilde{U},\,\widetilde{V}]=T(U_t-V_x+[U,\, V])T^{-1}=0,$
which yields the same equation (\ref{cmkdv}) with $q(x,t)\rightarrow \widetilde{q}(x,t)$, that is, $\widetilde{q}$ in the spectral problem (\ref{nlax3}) and (\ref{nlax4}) is also a solution of Eq.~(\ref{cmkdv}).

\subsection{Nonlocal Darboux matrix and generalized perturbation $(1, N-1)$-fold Darboux transformation}

To construct the nonlocal $(1, N-1)$-fold DT of Eq.~(\ref{cmkdv}), we consider the {\it nonlocal} Darboux matrix in Eq.~(\ref{dm}) in the form
\bee\nonumber
\hspace{-0.5in} T=T(\lambda)= \qquad\qquad \\
\left(\!\!\!\begin{array}{cc}
\lambda^{N}\!-\!\sum\limits_{j=0}^{N-1}\!\!A^{(j)}(x,t)\lambda^{j}    &-\sum\limits_{j=0}^{N-1}\!\!B^{(j)}(x,t)(-\lambda)^{j} \vspace{0.1in} \\
   -\!\sum\limits_{j=0}^{N-1}\!\!{B^{(j)}}^{*}(-x,t) (-\lambda)^{j}   &
   \sigma\!\! \left[\lambda^{N}\!-\!\sum\limits_{j=0}^{N-1}\!\!\!\!{A^{(j)}}^{*}\!(-x,t)\lambda^{j}\!\right]\end{array}
       \!\!\!\right) \quad \label{nlsm}
\ene
with the complex functions $A^{(j)}$ and $B^{(j)}\, (j=0,1,...,N-1)$ solving the linear algebraic system
$T(\lambda_k)\varphi_k(\lambda_k)=0\quad (k=1,2,...,N)$
 and the eigenfunctions $\varphi_i(\lambda_i)=(\phi(\lambda_i),\psi(\lambda_i))^T\, (i=1,2,...,N)$ are the solutions of the linear spectral problem (\ref{lax1}) and (\ref{lax2}) for the distinct $N$ spectral parameter $\lambda_i\, (i=1,2,...,N)$ and the same initial solution $q_0$. This case is not considered here. Notice that the entries in the second rows in matrix $T$ are all nonlocal functions, which differ from the local cases~\cite{wen1, wen2}.

To find the generalized (new) Darboux transformation, we here consider the nonlocal Darboux matrix (\ref{nlsm}) with only one spectral parameter $\lambda=\lambda_1$. Thus the condition $T(\lambda_1)\varphi(\lambda_1)=0$ leads to the linear algebraic system
 \bee \nonumber
 \left[\lambda_1^{N}\!-\!\sum\limits_{j=0}^{N-1}A^{(j)}(x,t)\lambda_1^{j}\right]\!\phi(\lambda_1, x, t) \qquad\quad\qquad\quad  \\
  -\sum\limits_{j=0}^{N-1}B^{(j)}(x,t) (-\lambda_1)^{j}\psi(\lambda_1, x, t)=0, \label{nlsag} \\
\nonumber \sigma \left[\lambda_1^{N*}\!-\!\sum\limits_{j=0}^{N-1}{A^{(j)}}(x,t)\lambda_1^{j*}\right]\!\psi^{*}(\lambda_1, -x ,t) \qquad\quad  \\
\label{nlsbg} -\!\sum\limits_{j=0}^{N-1}{B^{(j)}}(x,t) (-\lambda_1)^{j*}\phi^{*}(\lambda_1, -x, t)\!=\!0,
\ene
where $(\phi(\lambda_1), \psi(\lambda_1))^T$ is a solution of the Lax pair (\ref{lax1}) and (\ref{lax2}) with the spectral parameter $\lambda=\lambda_1$ and an initial solution $q_0(x,t)$.

We now know that the two linear algebraic equations (\ref{nlsag}) and (\ref{nlsbg}) contain the $2N$ unknown functions $A^{(j)}$ and $B^{(j)}\, (j=0,1,...,N-1)$. (i) if $N=1$, then we can determine only two complex functions $A^{(0)}$ and $B^{(0)}$ from Eqs.~(\ref{nlsag}) and (\ref{nlsbg}) such that we can not deduce the different functions $A^{(0)}$ and $B^{(0)}$ compared with ones from the usual DT; (ii) if $N>1$, then we have $2(N-1)>2$ free functions for $A^{(j)}$ and $B^{(j)}\, (j=0,1,...,N-1)$ in system (\ref{nlsag}) and (\ref{nlsbg}), which seems to be useful to determine the nonlocal Darboux matrix $T$, but it may be difficult to show the invariant conditions (\ref{t1}).

   To uniquely determine the functions $A^{(j)}$ and $B^{(j)}\, (j=0,1,...,N>1)$, we need to find more (e.g., $2(N-1)$) proper constraints for complex functions $A^{(j)}$ and $B^{(j)}$ except for the given constraints (\ref{nlsag}) and (\ref{nlsbg}). We now expand the expression
   \bee \nonumber
   T(\lambda_1)\varphi(\lambda_1)\big|_{\lambda_1=\lambda_1+\varepsilon}=
   \sum_{k=0}^{+ \infty}\sum\limits_{j=0}^{k}T^{(j)}(\lambda_1)\varphi^{(k-j)}(\lambda_1)\varepsilon^k
   \ene  at $\varepsilon=0$,
where $\varphi^{(k)}(\lambda_1)=\frac{1}{k!}\frac{\partial^k}{\partial \lambda^k}\varphi(\lambda)|_{\lambda=\lambda_1}$ and
$T^{(k)}(\lambda_1)=(T^{(k)}_{ij})_{2\times 2}$ with
\bee
T^{(k)}_{11}=C^{k}_N \lambda_1^{N\!-\!k}\!\!-\!\!\sum\limits_{j=k}^{N-1}\!C^{i}_j A^{(j)}(x,t)\lambda_1^{j-k},   \\ T^{(k)}_{12}=-\sum\limits_{j=k}^{N-1}C^{k}_j B^{(j)}(x,t)(-\lambda_1)^{j-k}, \vspace{0.1in} \\
T^{(k)}_{21}=-\sum\limits_{j=k}^{N-1}C^{k}_j {B^{(j)}}^{*}(-x,t)(-\lambda_1)^{j-k},    \\
T^{(k)}_{22}=\sigma\left[C^{k}_N \lambda_1^{N\!-\!k}\!\!-\!\!\sum\limits_{j=k}^{N-1}\!\!C^{k}_j {A^{(j)}}^{*}(-x,t)\lambda_1^{j-k}\right] \ene
with $C^{k}_j=\dfrac{j (j-1)...(j-k+1)}{k!}$\, ($k=0,1,...,N$).

To determine the $2N$ unknown functions $A^{(j)},\, B^{(j)}\,$ $(0\leq j\leq N-1)$ in Eq.~(\ref{nlsm}) with $\lambda=\lambda_1$, let
$\lim\limits_{\varepsilon \to 0}\dfrac{T(\lambda_1+\varepsilon) \varphi(\lambda_1+\varepsilon)}{\varepsilon^k}=0,\,\,\,
(k=0,1,...,N-1),$
which lead to the linear algebraic system with $2N$ equations
\bee\label{nls1sys}\begin{array}{r}
T^{(0)}(\lambda_1)\varphi^{(0)}(\lambda_1)=0,  \vspace{0.1in}\\
T^{(0)}(\lambda_1)\varphi^{(1)}(\lambda_1)+T^{(1)}(\lambda_1)\varphi^{(0)}(\lambda_1)=0, \vspace{0.1in} \\
\qquad \cdots\cdots,\qquad\qquad  \vspace{0.1in} \\
\sum\limits_{j=0}^{N-1}T^{(j)}(\lambda_1)\varphi^{(N-1-j)}(\lambda_1)=0,
\end{array}\ene
in which the first matrix system is just system ~(\ref{nlsag})-(\ref{nlsbg}).
Up to now, we have introduced  system (\ref{nls1sys}) containing $2N$ algebraic equations  with $2N$ unknowns functions $A^{(j)}$ and $B^{(j)}\, (j=0,1,...,N-1)$.
When the eigenvalue $\lambda_1$ is suitably chosen so that the determinant of coefficients for system~(\ref{nls1sys}) is nonzero, hence the Darboux matrix $T$ is uniquely determined by system~(\ref{nls1sys}).

\noindent {\bf Theorem 1.} {\it Let $\varphi(\lambda_1)=(\phi(\lambda_1),\psi(\lambda_1))^T$ be column vector solutions of the spectral problem (\ref{lax1})-(\ref{lax2}) for the spectral parameters $\lambda_1$ and initial solution $q_0(x,t)$ of Eq.~(\ref{cmkdv}), respectively, then the generalized perturbation $(1, N-1)$-fold Darboux transformation of Eq.~(\ref{cmkdv}) is defined by
\bee\label{sol}
 \widetilde{q}_{N}(x,t)=q_0(x,t)+2(-1)^{N-1} B^{(N)},
\ene
where $B^{(N)}=\frac{\Delta B^{(N)}}{\Delta_N}$  with $\Delta_N=$
\begin{widetext}
{\scriptsize{\[\left[\begin{array}{llllllll}
{\lambda_1}^{(N-1)} {\phi^{(0)}} & {\lambda_1}^{(N-2)} {\phi^{(0)}}  &\ldots & {\phi^{(0)}} & {\lambda_1}^{(N-1)} {\psi^{(0)}} & {\lambda_1}^{(N-2)} {\psi^{(0)}}  &\ldots & {\psi^{(0)}} \vspace{0.1in} \\
\Delta_{2,1}& \Delta_{2,2}  &\ldots &{\phi^{(1)}}  & \Delta_{2,N+1} &  \Delta_{2,N+2} &\ldots &{\psi^{(1)}} \vspace{0.1in}\\
 \ldots & \ldots      &\ldots           &\ldots &\ldots & \ldots      &\ldots           &\ldots  \vspace{0.1in} \\
\Delta_{N,1}  & \Delta_{N,2} &\ldots &{\phi^{(N-1)}} &\Delta_{N,N+1} & \Delta_{N,N+2} &\ldots &{\psi^{(N-1)}} \vspace{0.1in}\\
\sigma {\lambda_1}^{*(N-1)} {\psi^{(0)*}}(-x,t) & \sigma{\lambda_1}^{*(N-2)} {\psi^{(0)*}}(-x,t)   &\ldots & \sigma{\psi^{(0)*}}(-x,t)  & {(-\lambda_1^*)}^{(N-1)}{\phi^{(0)*}}(-x,t) &{(-\lambda_1^*)}^{(N-2)} {\phi^{(0)*}}(-x,t)  &\ldots & {\phi^{(0)*}}(-x,t)\vspace{0.1in} \\
\sigma\Delta_{N+2,1} & \sigma\Delta_{N+2,2} &\ldots & \sigma {\psi^{(1)*}}(-x,t)  &(-1)^{(N-1)}\Delta_{N+2,N+1} & (-1)^{(N-2)}\Delta_{N+2,N+2}  &\ldots &{\phi^{(1)}}(-x,t) \vspace{0.1in}\\
 \ldots & \ldots      &\ldots           &\ldots &\ldots & \ldots      &\ldots           &\ldots  \\
\sigma\Delta_{2N,1}  & \sigma\Delta_{2N,2}  &\ldots &\sigma{\psi^{(N-1)*}}(-x,t) &(-1)^{(N-1)}\Delta_{2N,N+1}^{(i)} & (-1)^{(N-2)}\Delta_{2N,N+2}  &\ldots &{\phi^{(N-1)}}(-x,t) \\
\end{array}\right],
\]}}
where $\Delta_{j,s}\, (1\leq j, s\leq 2N$) are given by the following formulae:
\bee\nonumber
\Delta_{j,s}=\begin{cases}
 \sum\limits_{k=0}^{j-1}C^{k}_{N-s} {\lambda_1}^{(N-s-k)} {\phi^{(j-1-k)}}
\quad {\rm for} \quad 1\leq j, s\leq N, \vspace{0.1in} \\
\sum\limits_{k=0}^{j-1}C^{k}_{2N-s} {\lambda_1}^{(2N-s-k)} {\psi^{(j-1-k)}}\quad {\rm for} \quad 1\leq j\leq N,\, N+1\leq s\leq 2N, \vspace{0.1in}\\
\sigma\sum\limits_{k=0}^{j-(N+1)}C^{k}_{N-s} {\lambda_1}^{*(N-s-k)} {\psi^{(j-N-1-k)*}}(-x,t) \quad {\rm for} \quad
N+1\leq j\leq 2N,\, 1\leq s\leq N, \vspace{0.1in}\\
\sum\limits_{k=0}^{j-(N+1)}C^{k}_{2N-s} {\lambda_1}^{*(2N-s-k)} {\phi^{(j-N-1-k)*}}(-x,t) \quad {\rm for} \quad
 N+1\leq j, s\leq 2N \end{cases} \ene
and $\Delta {B^{(N)}}$ is determined using the determinant $\Delta_N$ by replacing its
$(N+1)$-th column with the column vector $b=(b_j)_{2N\times 1}$, where
\bee\nonumber
b_j=\begin{cases} \sum\limits_{k=0}^{j-1}C^{k}_{N} {\lambda_1}^{(N-k)} {\phi^{(j-1-k)}}  \quad {\rm for} \quad 1\leq j\leq N, \vspace{0.1in} \\ \sigma \sum\limits_{k=0}^{j-(N+1)}C^{k}_{N} {\lambda_1}^{*(N-k)} {\psi^{(j-N-1-k)*}}(-x,t)  \quad {\rm for} \quad  N+1\leq j\leq 2N.
\end{cases} \qquad\qquad \ene
\end{widetext} }

\subsection{Generalized perturbation $(n, M=N-n)$-fold Darboux transformation}

We further extend the above-obtained nonlocal $(1, N-1)$-fold DT of Eq.~(\ref{cmkdv}), in which we use only one spectral parameter $\lambda=\lambda_1$ and the $m_1$th-order derivatives of $T(\lambda_1)$ and $\varphi(\lambda_1)$ with $m_1=1,2,...,N-1$. Nowadays we use $n$ ($1\leq n<N$) distinct spectral parameters $\lambda_i\, (i=1,2,...,n)$ and their corresponding highest order $m_i$ ($i=1,2,...,n)$  derivatives, where these non-negative integers $n,\, m_i$ are required to satisfy $N=n+\sum_{i=1}^nm_i=n+M$ with $M=\sum_{i=1}^nm_i$, where $N$ is the same as one in the Darboux matrix $T(\lambda)$ (\ref{nlsm}).

Similarly, we still consider the Darboux matrix (\ref{nlsm}) and the eigenfunctions $\varphi_i(\lambda_i)\, (i=1,2,...,n)$ are the solutions of the linear spectral problem (\ref{lax1}) and (\ref{lax2}) for the distinct spectral parameter $\lambda_i\, (i=1,2,...,n)$ and the same initial solution $q_0(x,t)$. Thus we have
\bee
\nonumber
 T(\lambda_i+\varepsilon) \varphi_i(\lambda_i+\varepsilon)
  \!\!=\!\!\sum_{k=0}^{+\infty}\sum\limits_{j=0}^{k}T^{(j)}(\lambda_i)\varphi_i^{(k-j)}(\lambda_i)\varepsilon^k,
   \ene
 where $\varphi_i^{(k)}(\lambda_i)=\frac{1}{k!}\frac{\partial^k}{\partial \lambda_i^k}\varphi_i(\lambda)|_{\lambda=\lambda_i}\, (k=1,2,...,N)$. Let $\lim\limits_{\varepsilon \to 0}\dfrac{T(\lambda_i+\varepsilon) \varphi_i(\lambda_i+\varepsilon)}{\varepsilon^{k}}=0$
with $i=1,2,...,n$ and $k=0,1,...,m_i$ that we obtain the linear algebraic system with the $2N$ equations ($N=n+\sum_{k=1}^nm_k$):
\bee\label{nlssys2}\begin{array}{r}
T^{(0)}(\lambda_i)\varphi_i^{(0)}(\lambda_i)=0,  \vspace{0.1in}\\
T^{(0)}(\lambda_i)\varphi_i^{(1)}(\lambda_i)+T^{(1)}(\lambda_i)\varphi_i^{(0)}(\lambda_i)=0, \vspace{0.1in} \\
\qquad \cdots\cdots,\qquad\qquad  \vspace{0.1in} \\
\sum\limits_{j=0}^{m_i}T^{(j)}(\lambda_i)\varphi_i^{(m_i-j)}(\lambda_i)=0,
\end{array}\ene
$i=1,2,...,n$, in which we know that some first systems for every index $i$, i.e., $T^{(0)}(\lambda_i)\varphi^{(0)}(\lambda_i)=T(\lambda_i)\varphi(\lambda_i)=0$ with $i=1$ are just some ones in Eqs.~(\ref{nlsag}) and (\ref{nlsbg}), but they are different if there exist at least one index $m_i\not=0$. For the chosen spectral parameters $\lambda_i\, (i=1,2,...,n)$, the derterminant of coefficients of system (\ref{nlssys2}) for the $2N$ variables $A^{(j)},\, B^{(j)}$ are non-zero such that we can determine them by using the Cramer's rule.

\vspace{0.1in}
\noindent {\bf Theorem 2.} {\it Let $\varphi_i(\lambda_i)=(\phi_i(\lambda_i),\psi_i(\lambda_i))^T\, (i=1,2,...,n)$ be column vector solutions of Lax pair (\ref{lax1}) and (\ref{lax2}) for the spectral parameters $\lambda_i\, (i=1,2,...,n)$ and initial solution $q_0(x,t)$ of Eq.~(\ref{cmkdv}), respectively, then the generalized perturbation $(n, M)$-fold Darboux transformation of Eq.~(\ref{cmkdv}) is given by
\bee\label{sol}
 \widetilde{q}_{N}(x,t)=q_0(x,t)+2(-1)^{N-1} B^{(N)},
\ene
where $B^{(N)}=\frac{\Delta B^{(N)}}{\Delta_N}$ ($N=n+\sum_{i=1}^nm_i)$  with $\Delta_N={\rm det}([\Delta^{(1)}_{m_1+1}\cdots\Delta^{(n)}_{m_n+1}]^T)$
and $\Delta^{(i)}_{m_i+1}\!=\!(\Delta^{(i)}_{j,s})_{2(m_i+1)\times 2N}\!$
\begin{widetext}
{\scriptsize{\[=\!\left[\begin{array}{llllllll}
{\lambda_i}^{(N-1)} {\phi_i^{(0)}} & {\lambda_i}^{(N-2)} {\phi_i^{(0)}}  &\ldots & {\phi_i^{(0)}} & {\lambda_i}^{(N-1)} {\psi_i^{(0)}} & {\lambda_i}^{(N-2)} {\psi_i^{(0)}}  &\ldots & {\psi_i^{(0)}} \vspace{0.1in} \\
\Delta_{2,1}^{(i)}& \Delta_{2,2}^{(i)}  &\ldots &{\phi_i^{(1)}}  & \Delta_{2,N+1}^{(i)} &  \Delta_{2,N+2}^{(i)} &\ldots &{\psi_i^{(1)}} \vspace{0.1in}\\
 \ldots & \ldots      &\ldots           &\ldots &\ldots & \ldots      &\ldots           &\ldots  \vspace{0.1in} \\
\Delta_{m_i+1,1}^{(i)}  & \Delta_{m_i+1,2}^{(i)} &\ldots &{\phi_i^{(m_i)}} &\Delta_{m_i+1,N+1}^{(i)}  & \Delta_{m_i+1,N+2}^{(i)}  &\ldots &{\psi_i^{(m_i)}} \vspace{0.1in}\\
\sigma {\lambda_i}^{*(N-1)} {\psi_i^{(0)*}}(-x,t) & \sigma{\lambda_i}^{*(N-2)} {\psi_i^{(0)*}}(-x,t)   &\ldots & \sigma{\psi_i^{(0)*}}(-x,t)  & {(-\lambda_i^*)}^{(N-1)}{\phi_i^{(0)*}}(-x,t) &{(-\lambda_i^*)}^{(N-2)} {\phi_i^{(0)*}}(-x,t)  &\ldots & {\phi_i^{(0)*}}(-x,t)\vspace{0.1in} \\
\sigma\Delta_{m_i+3,1}^{(i)} & \sigma\Delta_{m_i+3,2}^{(i)}  &\ldots & \sigma {\psi_i^{(1)*}}(-x,t)  &(-1)^{(N-1)}\Delta_{m_i+3,N+1}^{(i)} & (-1)^{(N-2)}\Delta_{m_i+3,N+2}^{(i)}  &\ldots &{\phi_i^{(1)}}(-x,t) \vspace{0.1in}\\
 \ldots & \ldots      &\ldots           &\ldots &\ldots & \ldots      &\ldots           &\ldots  \\
\sigma\Delta_{2(m_i+1),1}^{(i)}  & \sigma\Delta_{2(m_i+1),2}^{(i)}  &\ldots &{\psi_i^{(m_i)*}}(-x,t) &(-1)^{(N-1)}\Delta_{2(m_i+1),N+1}^{(i)} & (-1)^{(N-2)}\Delta_{2(m_i+1),N+2}^{(i)}  &\ldots &{\phi_i^{(m_i)}}(-x,t) \\
\end{array}\right],
\]}}
where $\Delta_{j,s}^{(i)}\, (1\leq i\leq n, 1\leq j\leq 2(m_i+1), 1\leq s\leq 2N$) are given by the following formulae:\\
\bee
\Delta_{j,s}^{(i)}=\begin{cases}
 \sum\limits_{k=0}^{j-1}C^{k}_{N-s} {\lambda_i}^{(N-s-k)} {\phi_i^{(j-1-k)}}
\quad {\rm for} \quad 1\leq j\leq m_i+1,\, 1\leq s\leq N, \vspace{0.1in} \\
\sum\limits_{k=0}^{j-1}C^{k}_{2N-s} {\lambda_i}^{(2N-s-k)} {\psi_i^{(j-1-k)}}\quad {\rm for} \quad 1\leq j\leq m_i+1,\, N+1\leq s\leq 2N, \vspace{0.1in}\\
\sigma\sum\limits_{k=0}^{j-(N+1)}C^{k}_{N-s} {\lambda_i}^{*(N-s-k)} {\psi_i^{(j-N-1-k)*}}(-x,t) \quad {\rm for} \quad
m_i+2\leq j\leq 2(m_i+1),\, 1\leq s\leq N, \vspace{0.1in}\\
\sum\limits_{k=0}^{j-(N+1)}C^{k}_{2N-s} {\lambda_i}^{*(2N-s-k)} {\phi_i^{(j-N-1-k)*}}(-x,t) \quad {\rm for} \quad
m_i+2\leq j\leq 2(m_i+1),\, N+1\leq s\leq 2N \end{cases} \ene
and $\Delta {B^{(N)}}$ is given by the determinant $\Delta_N$ by replacing its
$(N+1)$-th column with the column vector $(b^{(1)}\cdots b^{(n)})^T$, where $b^{(i)}=(b_j^{(i)})_{2(m_i+1)\times 1}\ (1\leq i\leq n), $ in which
\bee
b_j^{(i)}=\begin{cases} \sum\limits_{k=0}^{j-1}C^{k}_{N} {\lambda_i}^{(N-k)} {\phi_i^{(j-1-k)}}  \quad {\rm for} \quad 1\leq j\leq m_i+1, \vspace{0.1in} \\ \sigma \sum\limits_{k=0}^{j-(N+1)}C^{k}_{N} {\lambda_i}^{*(N-k)} {\psi_i^{(j-N-1-k)*}}(-x,t)  \quad {\rm for} \quad  m_i+2\leq j\leq 2(m_i+1).
\end{cases} \qquad\qquad \ene
\end{widetext} }

\noindent {\bf Remark.}\, (i) if the number of  the spectral parameters is one, that is, $n=1$, in which $m_1=N-1$, then  Theorem 2 reduces to Theorem 1; (ii) if $n=N$ and $m_i=0 \, (i=1,2,..,n)$, then  Theorem 2 reduces to the usual $N$-fold DT~\cite{lm}, that is, the generalized perturbation $(n, M)$-fold DT is a new extension of the usual $N$-fold DT. Thus the generalized perturbation $(n, M)$-fold DT can be used to obtain not only solitons (which are similar to ones obtained by using  the usual $N$-fold DT) but also new solutions including higher-order rational solutions (see Sec.IID).

\subsection{Higher-order rational soliton structures and dynamical behaviors}

 We here study the rational solitons of Eq.~(\ref{cmkdv}) by using the obtained nonlocal $(1, N-1)$-fold DT in Theorem 1.
 We start from the  general `seed' plane-wave solution of Eq.~(\ref{cmkdv}) in the form $q_0(x,t)=\rho e^{-i\sigma \rho^2 t+i\delta}$, where $\rho\not=0,\, \delta$ are both real-valued constant. Without loss of generality, we set $\delta=0$ and substitute the initial plane-wave solution $q_0(x,t)$ into the Lax pair (\ref{lax1}) and (\ref{lax2}), we can give its the solution of Eqs.~(\ref{lax1}) and (\ref{lax2}) with the  spectral parameter $\lambda=\lambda_{1}$ as follows:
\bee\nonumber
\varphi=\!\!\left[\!\!
\begin{array}{c}
(C_1 e^{A}+C_2 e^{-A})e^{-i\sigma \rho^2 t/2} \vspace{0.1in}\\
\left(C_1\frac{\sqrt{\lambda_1^2-\sigma \rho^2}-\lambda_1}{\rho} e^{A}-C_2 \frac{\sqrt{\lambda_1^2-\sigma \rho^2}+\lambda_1}{\rho} e^{-A}\right)e^{i\sigma \rho^2 t/2}
\end{array}
\right],\\\label{nls1v1}
\ene
where $A=\sqrt{\lambda_1^2-\sigma \rho^2}\left[x-i\lambda_1 t+\Theta(\varepsilon)\right]$ with $
\Theta(\varepsilon)=\sum\limits_{k=1}^{N}(b_k+ic_k)\varepsilon^{2k}$, $C_{1,2}$, $b_k, c_k (k=1,2,...,N)$ are any real-valued parameters, and $\varepsilon$ is a small parameter.

 Here we consider the defocusing case $\sigma=-1$ in Eq.~(\ref{cmkdv}). We choose $\lambda_1=i \rho+\varepsilon^2$ and expand the vector function $\varphi(\varepsilon^2)$ in Eq.~(\ref{nls1v1}) at $\varepsilon=0$ such that we obtain $\varphi(\varepsilon^2)=\varphi^{(0)}+\varphi^{(1)}\varepsilon^2+\varphi^{(2)}\varepsilon^4+\varphi^{(3)}\varepsilon^6+\cdots$,
where
\bee\nonumber
\varphi^{(0)}=\left(\begin{array}{c} \phi^{(0)}\\ \psi^{(0)} \end{array}\right)
=(C_1+C_2)\left[\begin{array}{c}  e^{\frac{i\rho^2 t}{2}} \vspace{0.05in}\\ -ie^{-\frac{i\rho^2 t}{2}} \end{array}  \label{nlsp0}
\right], \ene
\bee\nonumber
\begin{array}{rl}
\varphi^{(1)}=(C_1+C_2)\left[\begin{array}{c} i\rho (x+\rho t)^2e^{\frac{i\rho^2 t}{2}} \vspace{0.1in}\\
Q e^{-\frac{i\rho^2 t}{2}}  \end{array}  \label{nlsp1}
\right] \qquad
\end{array}\ene
with $Q=\rho^2(x^2+\rho^2 t^2)+2\rho^3xt+2i\rho(x+\rho t)-1$ and  $(\phi^{(i)},\psi^{(i)})^T (i=2,3,...)$  are omitted here. Notice that we find that $(\phi^{(i)},\psi^{(i)})^T$ have the same factor $C_1+C_2$, which does not affect the solutions of Eq.~(\ref{cmkdv}) (see Eq.~(\ref{sol})).

According to the generalized perturbation $(1, N-1)$-fold DT in Theorem 1, we will discuss the following four cases with $N =1,2,3,4$. In particular, for $N=1$, we obtain a trivial plane wave solution in terms of the $(1,0)$-fold DT $\widetilde{q}_1\!=\!\rho e^{-i\sigma \rho^2 t+i\delta}+\!2 {{B^{(1)}}}=-q_0.$

\vspace{0.1in}
 {\bf  Case I. \, First-order rational soliton structures and dynamical behaviors.}
   When $N=2$, based on the generalized perturbation $(1,1)$-fold DT in Theorem 1, we derive the regular first-order rational solution of Eq.~(\ref{cmkdv})
\bee\label{nnlss10}
 \widetilde{q}_{2g}(x,t)=\dfrac{2\rho^3(x^2-\rho^2t^2)+\rho+2i\rho^2(x+2\rho t)}{2\rho^2(x^2-\rho^2t^2)-1+2i\rho x}e^{i\rho^2t},\quad
   \ene
For any non-zero real parameter $\rho$, the solution $\widetilde{q}_{2g}(x,t)$ is an either even nor odd function about space. i.e., $\widetilde{q}_{2g}(-x,t)\not=\pm\widetilde{q}_{2g}(x,t)$ Therefore, it can not reduce to the solution of the corresponding NLS equation (\ref{nls1}).

\begin{figure}[!t]
\begin{center}
\vspace{0.05in}{\scalebox{0.35}[0.35]{\includegraphics{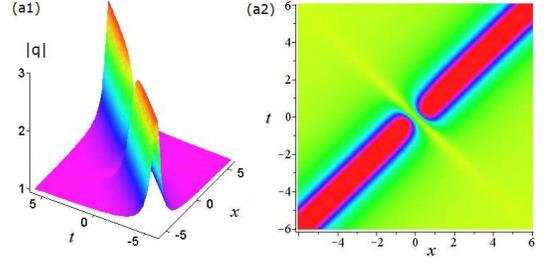}}}
\end{center}
	\vspace{-0.2in} \caption{\small (color online). (a1) the wave profile $|\widetilde{q}_{2}(x,t)|$ and (a2) density profile of the first-order rational soliton solution (\ref{nnlss1}).} \label{fig1}
\end{figure}

For convenience, we choose $\rho=1$ such that the regular rational solution  of Eq.~(\ref{cmkdv}) is given by
 \bee\label{nnlss1}
 \widetilde{q}_2(x,t)=\dfrac{2(x^2-t^2)+1+2i(x+2t)}{2(x^2-t^2)-1+2ix}e^{it},
   \ene
whose profile is displayed is Fig.~\ref{fig1}, which differs from the usual solitons~\cite{nls0} and may be regarded as a combination of the bright and dark solitons. We find that $\widetilde{q}_2^{*}(-x,t)\not=\pm\widetilde{q}_2^{*}(x,t)$, that is, the rational soliton (\ref{nnlss1}) does not satisfy Eq.~(\ref{nls1}).

 To understand the wave structures of the first-order rational soliton (\ref{nnlss1}), we firstly consider its intensity:
 \bee \label{am}
 |\widetilde{q}_{2}(x,t)|^2=\dfrac{[2(x^2-t^2)+1]^2+4(x+2t)^2}{[2(x^2-t^2)-1]^2+4x^2},\,\,\,
 \ene
 which has no singular point for all real $t, x$. We have $|\widetilde{q}_{2}(x,t)|^2\to 1$ as $x, t\to \infty$ with $x\not=t$, and $|\widetilde{q}_{2}(x,t)|^2\to 9$ as $x=t\to \infty$,  $|\widetilde{q}_{2}(x,t)|^2\to 1$ as $x=t\to 0$, $|\widetilde{q}_{2}(x,t)|^2=1$ as $x=-t$.

 We find that $ |\widetilde{q}_{2}(x,t)|^2$  has a family of critical points $(x,t)=(x,-x)$, that is, a family of real roots of system $\{\partial \left( |\widetilde{q}_2(x,t)|^{2}\right) /\partial t=0,\,\partial \left( |\widetilde{q}_2(x,t)|^{2}\right) /\partial x=0\}$. Moreover, the family of critical points can be shown to be minimum points (depressions) since
\bee
 \frac{\partial ^{2}\left( |\widetilde{q}_{2}(x,t)|^{2}\right) }{\partial t^{2}}=\frac{\partial ^{2}\left( |\widetilde{q}_{2}(x,t)|^{2}\right) }{\partial x^{2}}=\frac{16}{4x^2+1}>0
 \nonumber
 \ene and
\bee
\nonumber
\left[\!\frac{\partial ^{2}\left( |\widetilde{q}_{2}(x,t)|^{2}\right) }{\partial x\partial t}\!\right] ^{2}
\!-\!\frac{\partial ^{2}\left(\!|\widetilde{q}_{2}(x,t)|^{2}\right) }{\partial t^{2}}\frac{\partial ^{2}
\left( |\widetilde{q}_{2}(x,t)|^{2}\right) }{\partial x^{2}}=0
\ene at the family of critical points $(x, t)=(x, -x)$.

\begin{figure}[!t]
\begin{center}
\vspace{0.05in}{\scalebox{0.4}[0.4]{\includegraphics{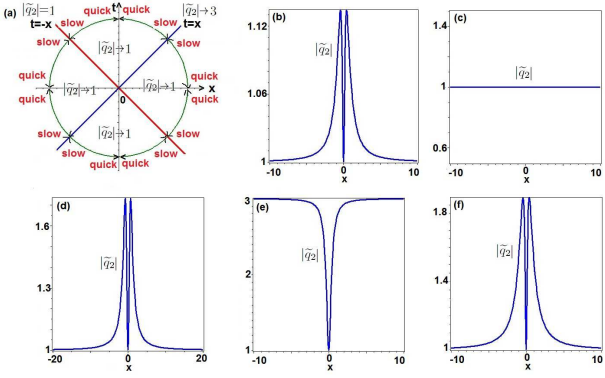}}}
\end{center}
	\vspace{-0.2in} \caption{\small (color online). (a) the limits of first-order rational soliton (\ref{nnlss1}) in $(x, t)$-space. The
amplitude wave profiles of solution  (\ref{nnlss1}) for: (b) $t=-2$, (c) $t=-x$, (d) $t=0$, (e) $t=x$, and (f) $t=2$. Here `quick' and `slow' imply that the amplitude of the solution (\ref{nnlss1}) quickly and slowly approaches to $1$ along the arrow directions, respectively.} \label{fig1-point}
\end{figure}

We now check the profiles of $|\widetilde{q}_{2}(x,t)|$ from the different lines $t=\alpha x$ with  $\alpha\in \mathbb{R}$. There are two critical lines $t=\pm x$. For the case $|\alpha|>1$, $|\widetilde{q}_{2}(x,t)|\to 1$ as $x\to \infty$ and $|\widetilde{q}_{2}(x,t)|$ more quickly approaches to $1$ as $|\alpha|$ increases. For the case $0\leq |\alpha|<1$, $|\widetilde{q}_{2}(x,t)|\to 1$ as $x\to \infty$ and $|\widetilde{q}_{2}(x,t)|$ more quickly approaches to $1$ as $|\alpha|$ decreases. For the case $\alpha=-1$, we have $|\widetilde{q}_{2}(x,t)|=1$ for any $x$, however  for the case $\alpha=1$, we have $|\widetilde{q}_{2}(x,t)|\to 3$ as $x\to\infty$ (see Fig.~\ref{fig1-point}).

\begin{figure}[!t]
	\begin{center}
	\vspace{0.05in}	{\scalebox{0.4}[0.4]{\includegraphics{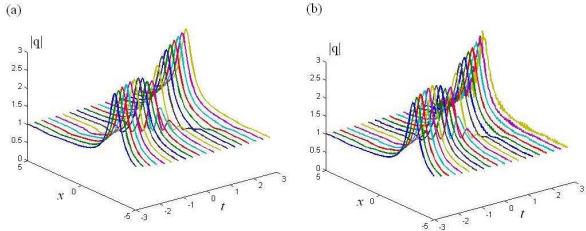}}}
	\end{center}
	\vspace{-0.2in} \caption{\small (color online).  The first-order rational soliton $\widetilde{q}_{2}$ given by Eq.~(\ref{nnlss1}). (a) exact solution, (b) time evolution of the wave using the exact solution (\ref{nnlss1}) perturbated by a $1\%$ noise as the initial condition. The parameters are $\sigma=-1, \rho=1$.} \label{1-rw-nnls}
\end{figure}

To further illustrate the wave propagations of the first-order rational soliton  (\ref{nnlss1}), we here consider its
dynamical behaviors by comparing itself with its time evolution using it as the initial condition with a small noise in terms of numerical simulations.
Fig.~\ref{1-rw-nnls}a exhibits the exact first-order rational soliton (\ref{nnlss1}) of Eq.~(\ref{cmkdv}). Fig.~\ref{1-rw-nnls}b displays the time evolution of rational soliton (\ref{nnlss1}) perturbated by a small noise $1\%$ as the initial condition. Fig.~\ref{1-rw-nnls}b implies the almost stable propagation, except for some oscillations as time approaches to $3$.

\vspace{0.2in}
{\bf Case II.\, Second-order rational solitons and dynamical behaviors.}
  When $N=3$,  based on the generalized perturbation $(1,2)$-fold DT, we can derive the second-order rational soliton of Eq.~(\ref{cmkdv}) with $\rho=1, C_1=1, C_2=2$ and other two free parameters $b_1,c_1$ as below:
\begin{eqnarray}
\widetilde{q}_3(x,t)=\!\rho e^{-i\sigma \rho^2 t+i\delta}-\!2 {{B^{(3)}}}\!=\!-\frac{G(x,t)}{F(x,t)}e^{it}, \quad  \label{nnls2}
 \end{eqnarray}
 where we have introduced the functions $F(x,t)$ and $G(x,t)$ as
\bee\nonumber\begin{array}{rl}
 G(x,t)=&\!\!144 i  x t^2 b_1- 144 i  x b_1+72 b_1+ 144 i  x t c_1 \vspace{0.05in}\\
        &\!\!-72 x^2-144 x t+216 t^2+144 c_1 x-72 c_1 t \vspace{0.05in}\\
        &\!\! -24 x^4+120 t^4-192 x^3 t-288 x^2 t^2+192 x t^3  \vspace{0.05in}\\
        &\!\! -72 x^2 b_1 -72 t^2 b_1+288 x t b_1-36 b_1^2+48 c_1 t^3 \vspace{0.05in}\\
        &\!\!-48 x^4 t^2+48 t^4 x^2+16 x^6-16 t^6+144 c_1 x^2 t \vspace{0.05in}\\
        &\!\!-36 c_1^2+ 72 i  c_1+ 48 i  x^5- 144 i  x^2 c_1- 96 i  x^3 t^2 \vspace{0.05in}\\
        &\!\!- 144 i  t- 36 i x- 192 i  x^2 t^3- 360 i  t^2 x+ 96 i  t^3 \vspace{0.05in}\\
        &\!\!+ 144 i  t b_1+ 48 i  x t^4+ 24 i  x^3+ 96 i  x^4 t+ 48 i  x^3 b_1 \vspace{0.05in}\\
        &\!\!+ 96 i  t^5- 144 i  t^2 c_1-9,
  \end{array} \ene
\bee\nonumber\begin{array}{rl}
 F(x,t)=&\!\! 144 i x t c_1+72 x^2-72 t^2+72 c_1 t-72 x^4 \vspace{0.05in}\\
        &\!\!-120 t^4-72 x^2 b_1-72 t^2 b_1-36 b_1^2+48 c_1 t^3 \vspace{0.05in}\\
        &\!\!-48 x^4 t^2+48 t^4 x^2+16 x^6-16 t^6+144 c_1 x^2 t \vspace{0.05in}\\
        &\!\!-36 c_1^2+ 144 i  x t^2 b_1+ 48 i  x t^4- 72 i  t^2 x- 72 i  x^3 \vspace{0.05in}\\
        &\!\!+ 48 i +x^5+ 36 i  x+ 48 i  x^3 b_1- 96 i  x^3 t^2-9.
 \end{array} \ene

\begin{figure}[t!]
	\begin{center}
{\scalebox{0.35}[0.35]{\includegraphics{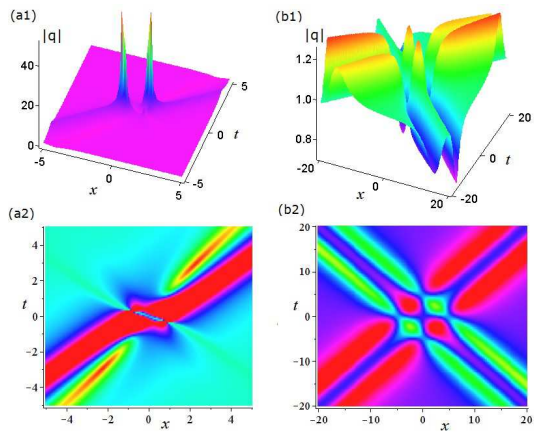}}}
	\end{center}
	\vspace{-0.25in} \caption{\small (color online).  The amplitude wave profile of the second-order rational soliton  $\widetilde{q}_3$ given by Eq.~(\ref{nnls2}) for the different parameters: (a1)-(a2)  $b_1=c_1=0$; (b1)-(b2) $b_1=100, c_1=0$.} \label{fig2}
\end{figure}

For the case $b_1=c_1=0$, the second-order rational soliton (\ref{nnls2})  exhibits the strong interaction (see Figs.~\ref{fig2}(a1)-(a2)), but for the case $b_1=100, c_1=0$, the second-order rational soliton (\ref{nnls2}) is split into the interactions of two bright solitons and two dark solitons (see Figs.~\ref{fig2}(b1)-(b2)).

\begin{figure}[ht!]
	\begin{center}
		{\scalebox{0.42}[0.42]{\includegraphics{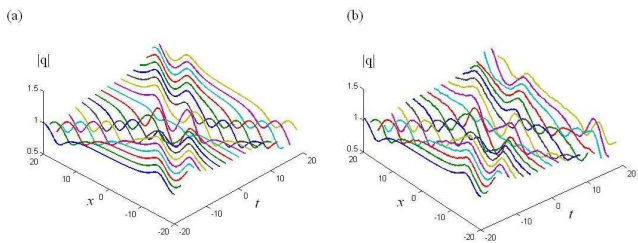}}}
	\end{center}
	\vspace{-0.25in} \caption{\small (color online).  The second-order interactive rational soliton $\widetilde{q}_{3}$ given by Eq.~(\ref{nnls2}) with  $b_1=100, c_1=0$ (see Figs.~\ref{fig2}(b1)-(b2)). (a) exact solution, (b) time evolution of the wave using the exact solution (\ref{nnls2}) perturbated by a $5\%$ noise as the initial condition.} \label{2-rw-nnls}
\end{figure}

Here we study the dynamical behaviors of solution (\ref{nnls2}) with $b_1=100, c_1=0$ (see Figs.~\ref{fig2}(b1)-(b2)).  Figs.~\ref{2-rw-nnls}a and \ref{2-rw-nnls}a  illustrate the exact second-order rational soliton (\ref{nnls2}) and its time evolution  using its perturbation with a small noise $5\%$ as the initial condition, respectively. Particularly, Fig.~\ref{2-rw-nnls}b displays the almost stable wave propagation of the second-order rational soliton (\ref{nnls2}).

\vspace{0.1in}
{\bf Case III.\,  Third-order rational solitons and dynamical behaviors.}
 When $N=4$, according to Theorem 1, we have ${B^{(4)}}=\frac{\Delta {B^{(4)}}}{\Delta_4}$ such that we obtain the third-order rational soliton of Eq.~(\ref{cmkdv}) in the form
 \bee \label{s4}
  \widetilde{q}_4(x,t)=\rho e^{-i\sigma \rho^2 t+i\delta}+\!2 {{B^{(4)}}},\ene
 in which we have chosen $\rho=1, C_1=1, C_2=2$ and leave four parameters $b_1,\, b_2, c_1, c_2$. The parameters $b_1,\, b_2, c_1, c_2$ excite the third-order rational solution to generate the abundant wave structures.

\begin{figure}[ht!]
	\begin{center}
	{\scalebox{0.43}[0.43]{\includegraphics{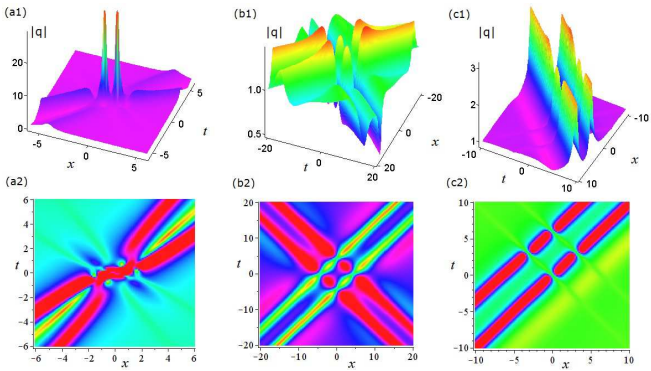}}}
	\end{center}
	\vspace{-0.2in} \caption{\small (color online).  The amplitude wave profile of the third-order rational soliton $\widetilde{q}_4$ given by Eq.~(\ref{s4}) for the different parameters: (a1)-(a2)  $b_{1,2}=c_{1,2}=0$; (b1)-(b2) $b_1=100, b_2=c_{1,2}=0$; (c1)-(c2) $b_2=1000, b_1=c_{1,2}=0$. } \label{fig3}
\end{figure}

\begin{itemize}
\item{} \, For the case  $b_{1,2}=c_{1,2}=0$, the third-order rational soliton  stays beside the origin in $(x,t)$-plane, that is, the third first-order rational solitons generate the strong interaction at the point $(x,t)=(0,0)$ (see Figs.~\ref{fig3}(a1)-(a2)).

\item{} \,  For  the case $b_1=100, b_2=c_{1,2}=0$ (i.e., at least one parameter is not zero in these parameters $\{b_1,\, b_2,\,c_1,\, c_2\}$),  the third-order rational soliton  can be split into the weak interaction among three solitons (see Figs.~\ref{fig3}(b1)-(b2)).

\item{}\,  For  the case $b_2=1000, b_1=c_{1,2}=0$, the third-order rational soliton  among three solitons are nearly split into three parallel solitons which include two bright solitons and one dark soliton (see Figs.~\ref{fig3}(c1)-(c2)).
\end{itemize}

\begin{figure}[!t]
	\begin{center}
		{\scalebox{0.42}[0.42]{\includegraphics{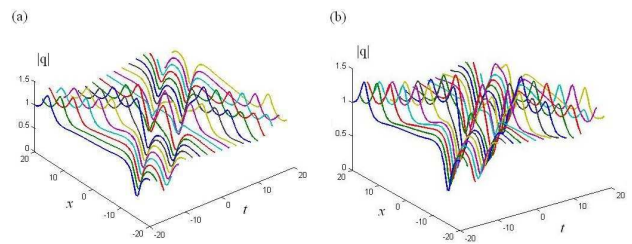}}}
	\end{center}
	\vspace{-0.25in} \caption{\small (color online).  The third-order interactive rational soliton $\widetilde{q}_{4}$  with $b_1=100, b_2=c_1=c_2=0$ . (a) exact solution, (b) time evolution of the wave using the exact solution $\widetilde{q}_{4}$ perturbated by a $1\%$ noise as the initial condition.} \label{n3-rw-b100}
\end{figure}

\begin{figure}[!t]
	\begin{center}
	{\scalebox{0.42}[0.42]{\includegraphics{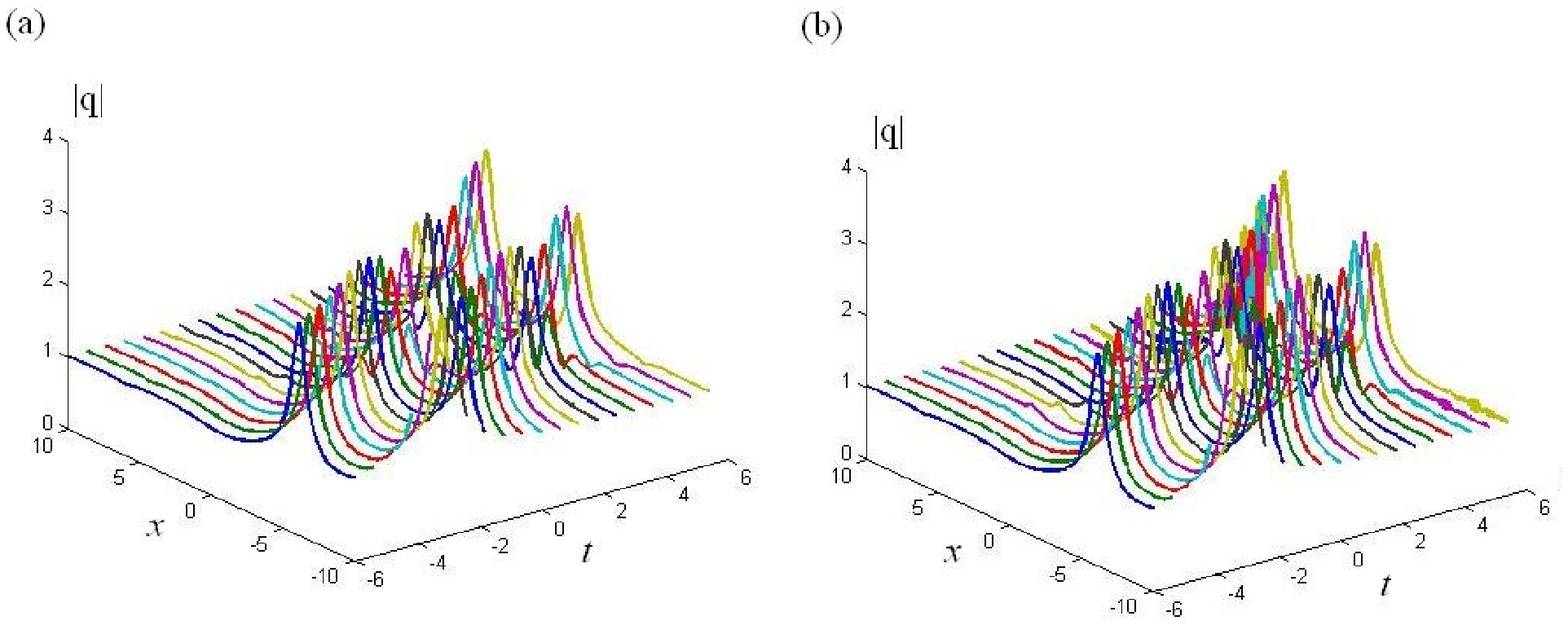}}}
	\end{center}
	\vspace{-0.25in} \caption{\small (color online).  The separatable third-order rational soliton $\widetilde{q}_{4}$ with $b_2=1000, b_1=c_1=c_2=0$. (a) exact solution, (b) time evolution of the wave using the exact solution  $\widetilde{q}_{4}$ perturbated by a $0.5\%$ noise as the initial condition.} \label{n3-rw-b21000}
\end{figure}

We here study the dynamical behaviors of solution $\widetilde{q}_{4}$ with $b_1=100, b_2=c_{1,2}=0$ (see Figs.~\ref{fig3}(b1)-(b2)) and $b_2=1000, b_1=c_{1,2}=0$ (see Figs.~\ref{fig3}(c1)-(c2)). Figs.~\ref{n3-rw-b100}-\ref{n3-rw-b21000} illustrate the exact third-order rational soliton  $\widetilde{q}_{4}$ of Eq.~(\ref{cmkdv}) and its time evolution using the exact solution $\widetilde{q}_{4}$ perturbated by a small noise (e.g., $1\%$ for Fig.~\ref{n3-rw-b100}b and $0.5\%$ for Fig.~\ref{n3-rw-b21000}b) as the initial conditions, respectively. Figs.~\ref{n3-rw-b100}b and \ref{n3-rw-b21000}b exhibit the almost stable wave propagations.

For the other cases $N>4$, we can also explicitly find the higher-order rational solitons of Eq.~(\ref{cmkdv}), which are omitted here and seem to also possess the interesting wave structures.

\section{Conclusions and discussions}

In conclusion, we have presented a novel method to construct the nonlocal version of generalized perturbation $(1, N-1)$-fold DT for the defocusing nonlocal NLS equation (\ref{cmkdv}) such that its higher-order rational solitons are found. The wave profiles of those rational solitons have been discussed in detailed for distinct parameters, which possess the bright and dark soliton structures.  Moreover, we also study the dynamical behaviors of these rational solutions with a small noise with the aid of numerical simulations such that some stable modes are found within some limit time. These results might be helpful for understanding physical phenomena described by Eq.~(\ref{cmkdv}) and finding possible application of rational solitons. We believe that the used idea in this paper is rather general and could also be extended to other single and vector nonlocal nonlinear wave models (e.g., ~\cite{yanaml15}), which will be presented in another literature.\vspace{0.1in}

  {\bf Acknowledgments} \vspace{0.1in}

 The authors would like to thank the referees for their valuable suggestions that have improved substantially
the paper. This work has been supported by the NSFC under Grant Nos. 11375030 and 11571346, the Beijing Natural Science Foundation under Grant No. 1153004, and China Postdoctoral Science Foundation under Grant No. 2015M570161.

\end{document}